\documentclass[aps,prl,twocolumn,showpacs,groupedaddress]{revtex4}
\usepackage{graphics}
\usepackage{multirow}
\usepackage{epsfig}

\begin{document}

\title{Precision measurement of the branching ratio in the $6\mathrm{P}_{3/2}$ decay of BaII with a single trapped ion}

\author{N. Kurz}
\email[]{nkurz129@u.washington.edu}
\homepage[Group homepage\hspace{5pt}]{http://faculty.washington.edu/blinov}
\author{M. R. Dietrich}
\author{Gang Shu}
\author{R. Bowler}
\author{J. Salacka}
\author{V. Mirgon}
\author{B. B. Blinov}
\affiliation{Department of Physics, University of Washington, Seattle, WA 98195}

\date{\today}

\begin{abstract}
We present a measurement of the branching ratios from the $6\mathrm{P}_{3/2}$ state of BaII into all dipole-allowed decay channels ($6\mathrm{S}_{1/2}$, $5\mathrm{D}_{3/2}$ and $5\mathrm{D}_{5/2}$).  Measurements were performed on single $^{138}\mathrm{Ba}^+$ ions in a linear Paul trap with a frequency-doubled mode-locked Ti:Sapphire laser resonant with the $6\mathrm{S}_{1/2}\rightarrow 6\mathrm{P}_{3/2}$ transition at 455 nm by detection of electron shelving into the dark $5\mathrm{D}_{5/2}$ state.  By driving a $\pi$ Rabi rotation with a single femtosecond pulse, an absolute measurement of the branching ratio to $5\mathrm{D}_{5/2}$ state was performed.  Combined with a measurement of the relative decay rates into $5\mathrm{D}_{3/2}$ and $5\mathrm{D}_{5/2}$ states performed with long trains of highly attenuated 455 nm pulses, it allowed the extraction of the absolute ratios of the other two decays.  Relative strengths normalized to unity are found to be 0.756$\pm$0.046, 0.0290$\pm$0.0015 and 0.215$\pm$0.0064 for $6\mathrm{S}_{1/2}$, $5\mathrm{D}_{3/2}$ and $5\mathrm{D}_{5/2}$ respectively.  This approximately constitutes a threefold improvement over the best previous measurements and is a sufficient level of precision to compare to calculated values for dipole matrix elements.
\end{abstract}

\pacs{31.15.bw, 32.70.Cs, 32.70.Fw}

\keywords{}

\maketitle

Single trapped ions are a valuable physical system for many applications including quantum computation \cite{CZ, Monroe}, frequency standards, optical metrology \cite{Berkeland}, precision searches for drifts in fundamental constants \cite{constants} and tests of exotic physical theories \cite{Fortson}.  Among the advantages over alternatives are their long trapping lifetimes and the relative ease of confining single ions to a small volume in a trap, thereby reducing the systematic effects and negating the need for quantum statistics.  The barium ion, particularly the odd 137 isotope with nuclear spin 3/2, has been proposed for use in quantum computation schemes with the hyperfine levels of the $6\mathrm{S}_{1/2}$ ground state for the qubit, as an optical frequency standard with a 2051 nm clock transition from $6\mathrm{S}_{1/2},\:\mathrm{F}=2,\:\mathrm{m}_\mathrm{F}=0$ to $5\mathrm{D}_{3/2},\:\mathrm{F}^\prime=0$, and as a test of parity-nonconservation with a small dipole coupling between the otherwise dipole-forbidden $6\mathrm{S}_{1/2}\rightarrow 5\mathrm{D}_{3/2}$ transition\cite{Fortson, Koerber1, Koerber2}.

Accurate models of atomic wave functions which include many-body interactions are necessary to calculate dipole and quadrupole matrix elements that appear in the calculations of transition rates, energy level shifts and line widths in the experiments mentioned above.  Measurements of branching ratios represent a better quantity from which to verify such values than, for example, precise measurements of the lifetimes of metastable states.  They are less prone to systematic uncertainties such as background gas quenching and stray fields to which the long waiting times (tens of seconds) required to accurately measure lifetimes are sensitive.  Here we present a single-ion measurement of the branching ratios from the $6\mathrm{P}_{3/2}$ state of $^{138}\mathrm{Ba}^+$ to the three states allowed via dipole transitions, $6\mathrm{S}_{1/2}$, $5\mathrm{D}_{3/2}$ and $5\mathrm{D}_{5/2}$.

A schematic of the optical and electronic arrangement of the experimental apparatus can be found in Fig. 1.  The ion trap itself is a linear Paul trap with radio-frequency quadrupole confining potential and DC voltage end caps in ultra high vacuum with operating pressures of about $10^{-11}$ torr.  The trap dimensions are $\sim\,$0.5 mm radially and $\sim\,$3.3 mm axially.  At $\sim\,$0.5 W of inductively-coupled RF power at $\sim\,$32 MHz and 100 V end cap potential, the trap secular frequencies are measured to be approximately $\omega_R/2\pi=5$ MHz and $\omega_Z/2\pi=2$ MHz.  A Xenon flash lamp is focussed into the trap to photoionize a thermal beam of atomic barium.  The barium ion has a lambda-type energy spectrum (shown in Fig. 2) with a strong $6\mathrm{S}_{1/2}\leftrightarrow 6\mathrm{P}_{1/2}$ transition at 493.4 nm which is used for both Doppler cooling and fluorescence detection, and a repump from the metastable $5\mathrm{D}_{3/2}$ state at 649.7 nm.  Both cooling and repumping are accomplished with commercial external-cavity diode lasers (Toptica TA-SHG 110 and DL 100 respectively).  The 493 nm light is generated by frequency doubling the output of a 986 nm diode with a $\mathrm{KNbO}_3$ crystal in a bow-tie enhancement cavity.  Frequency stabilization is accomplished via Doppler-free spectroscopy on vapor cells, Te$_2$ for the 493 nm laser and I$_2$ for the 650 nm laser.   Both lasers are matched onto a dichroic mirror and fiber-coupled for mode cleaning and to ensure collinearity at the trap.  A magnetic field of $\sim\,$4 G applied parallel to the cooling beams breaks the degeneracy of Zeeman sublevels.  Effective cooling is achieved with approximately 100 $\mu$W of blue light and 40 $\mu$W of red.  State discrimination is accomplished by observing ion fluorescence on the cooling transition collected with a Mitutoyo N.A.=0.28 objective with a Hamamatsu PMT behind a 493 nm interference filter.  Using a 100 ms detection interval, ``bright" and ``dark" histograms of photon counts are typically 30 and 200, respectively.  This is a sufficient separation to completely rule out the possibility of incorrect state determination.

Excitation from the ground state to the $6\mathrm{P}_{3/2}$ state is accomplished with broadband pulses with central wavelength of approximately 455 nm generated by frequency doubling a mode-locked Ti:Sapphire laser operating at 910 nm via single pass of a 5 mm $\beta\mathrm{BaB_2O_4}$ crystal.  Pulses had a duration of 400 fs before doubling as measured by intensity autocorrelation.  Pulse energy was adjusted by a DAC-controlled voltage variable RF attenuator (Minicircuits ZX73-2500) in line between the switching AOM and its driver.  For the experiment measuring the relative branching of the two D levels, a time-averaged power level of between 1.6 and 3.4 mW of 455 nm light was sufficient.  Single pulse selection for the Rabi rotation experiment was accomplished with the same AOM synchronized to the 76 MHz repetition rate of the laser using an intracavity-monitoring fast diode coupled to a Potato Semiconductor PO74G08A, 1.125 GHz bandwidth AND-gate, and an HP 8013B TTL Pulse Generator.  Extinction of neighboring pulses was better than 97\% at 455 nm, and asynchronicity was monitored with an HP 5335A universal counter and found to occur an average of 3-5\% of the time.  The maximum achievable single pulse energy was 1.68 nJ after transmission through the pulse selector.  Shot-to-shot variability of the pulse energy was approximately 10\% in the blue.  De-shelving from $5\mathrm{D}_{5/2}$ was accomplished with 50 ms flashes from a 614 nm high-power LED.

\begin{figure}[h]
\centering
\includegraphics[scale=0.4]{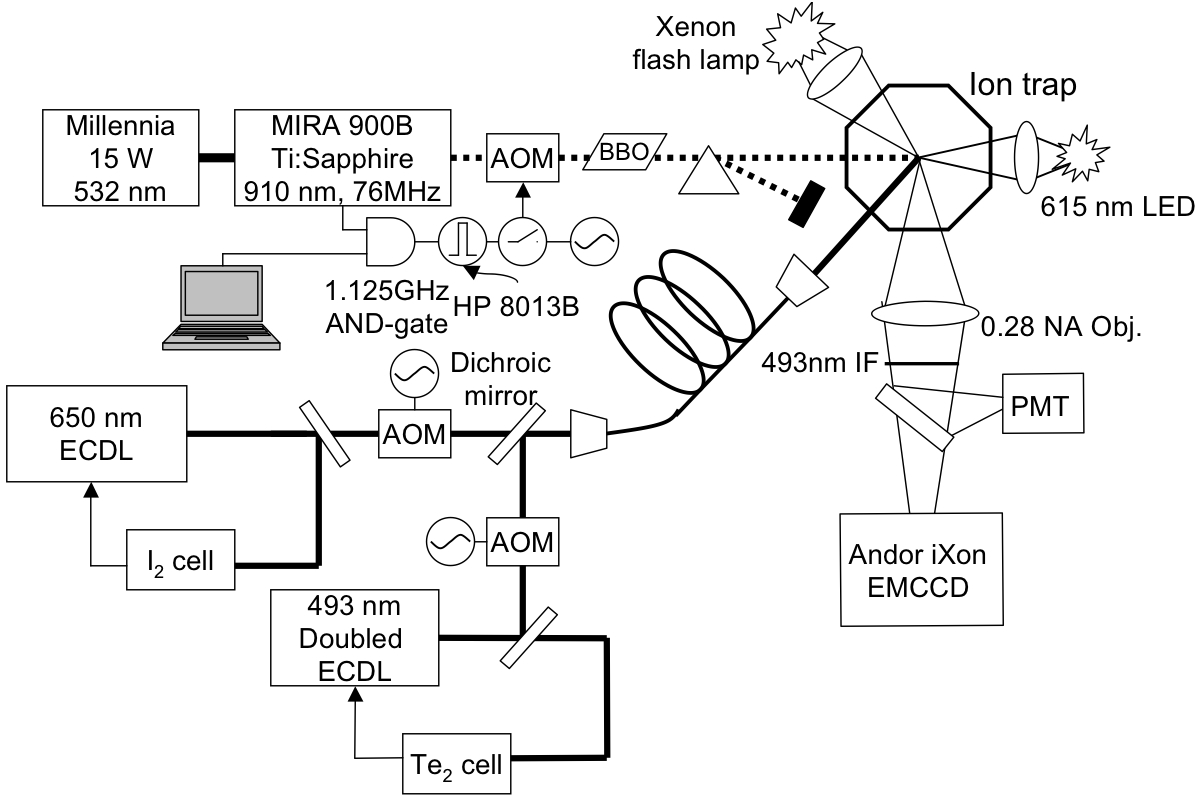}
\caption{Schematic overview of the apparatus used in this experiment.  After stabilization to their respective gas cells, cooling (493 nm) and repump lasers (650 nm) are coupled on a dichroic mirror and fiber coupled for mode-cleaning and to ensure collinearity.  Detection of ion fluorescence, after collection optics and filtering through a 493 nm narrow band interference filter, can be switched between a cooled electron-multiplied CCD (EMCCD) Andor iXon camera with single photon-sensitivity or a Hamamatsu photomultiplier tube (PMT).  The Xenon flash lamp has high spectral content in the UV and thereby photoionizes of a thermal beam of atomic barium.  The Ti:Sapphire, independently synchronized to the measurement apparatus through the 1.125 GHz gate and HP 8031B pulse generator, provides 400 fs pulses at 910 nm central wavelength which are doubled to 455 nm after a single pass through the BBO crystal.  The 614 nm high power LED de-shelves the ion from the $5\mathrm{D}_{5/2}$ state.}
\end{figure}
 
 \begin{figure}[h]
 \includegraphics[scale=0.4]{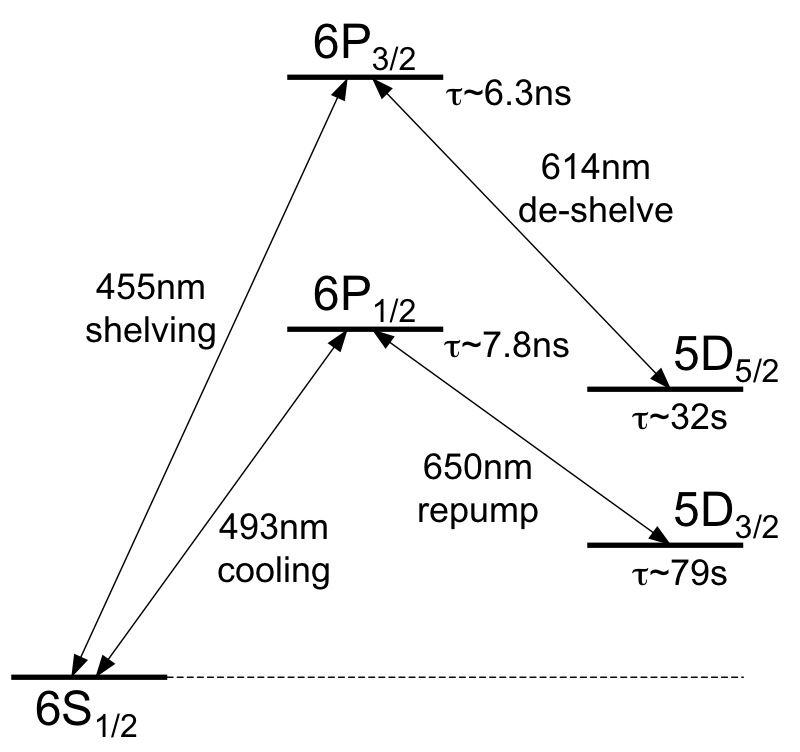}
 \caption{Relevant energy levels and lifetimes in the $^{138}\mathrm{Ba}^+$ ion.  D state lifetimes are experimental values \cite{Yu}.  The decays under consideration are from $6\mathrm{P}_{3/2}$ to $6\mathrm{S}_{1/2}$ at 455.403 nm, $5\mathrm{D}_{3/2}$ at 585.530 nm (not shown in digram) and $5\mathrm{D}_{5/2}$ at 614.171 nm.}
 \end{figure}

To establish a value for the relative strengths of the decays to the two D states, the ion was prepared in the $6\mathrm{S}_{1/2}$ state by optically pumping out of the $5\mathrm{D}_{3/2}$ state with the 650 nm laser.  With the two cooling lasers shuttered off, trains of highly attenuated 455 nm pulses of various duration from 0.1 to 10 ms and average powers of 1.6 to 3.4 mW were focused onto the ion, weakly exciting it to the $6\mathrm{P}_{3/2}$ state, which has a characteristic lifetime of 6.32 ns \cite{Pinnington}.  After this state decayed, the cooling lasers were turned on to look for ion fluorescence.  The absence of photons at 493 nm indicated that the ion was shelved to the $5\mathrm{D}_{5/2}$ state, but decays to $6\mathrm{S}_{1/2}$ and $5\mathrm{D}_{3/2}$ were indistinguishable using this method.  Plotting the probability of finding the ion in the dark state as a function of the exposure time to 455 nm light, one finds that the dark probability asymptotically approaches a constant value corresponding to the ratio of strengths of decay to the two D states as displayed in Fig. 3.  The solid line is a fit to the data using the expected form of this function, $P_{dark}=\Delta+B\left(1+e^{-\lambda t}\right)$ where $\Delta$ is any offset from zero due to false dark counts, mostly resulting from collisions of the ion with background gas particles and typically much less than 1\%, $B$ is the desired ratio and $\lambda$ is a parameter which depends on laser intensity and proximity to resonance.  Using this excitation and detection scheme, we derive the relative strength of D state decays from this asymptotic value of the dark state probability as a function of shelving pulse length.  Values for the ratios excluding decays to the ground state are 0.881 and 0.119 for $5\mathrm{D}_{5/2}$ and $5\mathrm{D}_{3/2}$ respectively.  Each has a 3\% statistical and 0.5\% systematic uncertainty, the latter arising mainly from leakage of cooling light during the shelving phase of the experiment and collisional de-excitation from the $5\mathrm{D}_{5/2}$ state due to background gas.  Optical attenuation of the cooling light was at least 85 dB, which ensures that the systematic decrease in the measured ratio is less than 0.5\%.  From extrapolated quenching rates at our operating pressure of $2.0\times10^{-11}$ torr over the 100 ms duration of each experiment, collisional de-excitation contributes to a systematically low value by of order 0.1\% \cite{Madej}.

\begin{figure}[h]
\centering
\includegraphics[scale=0.32]{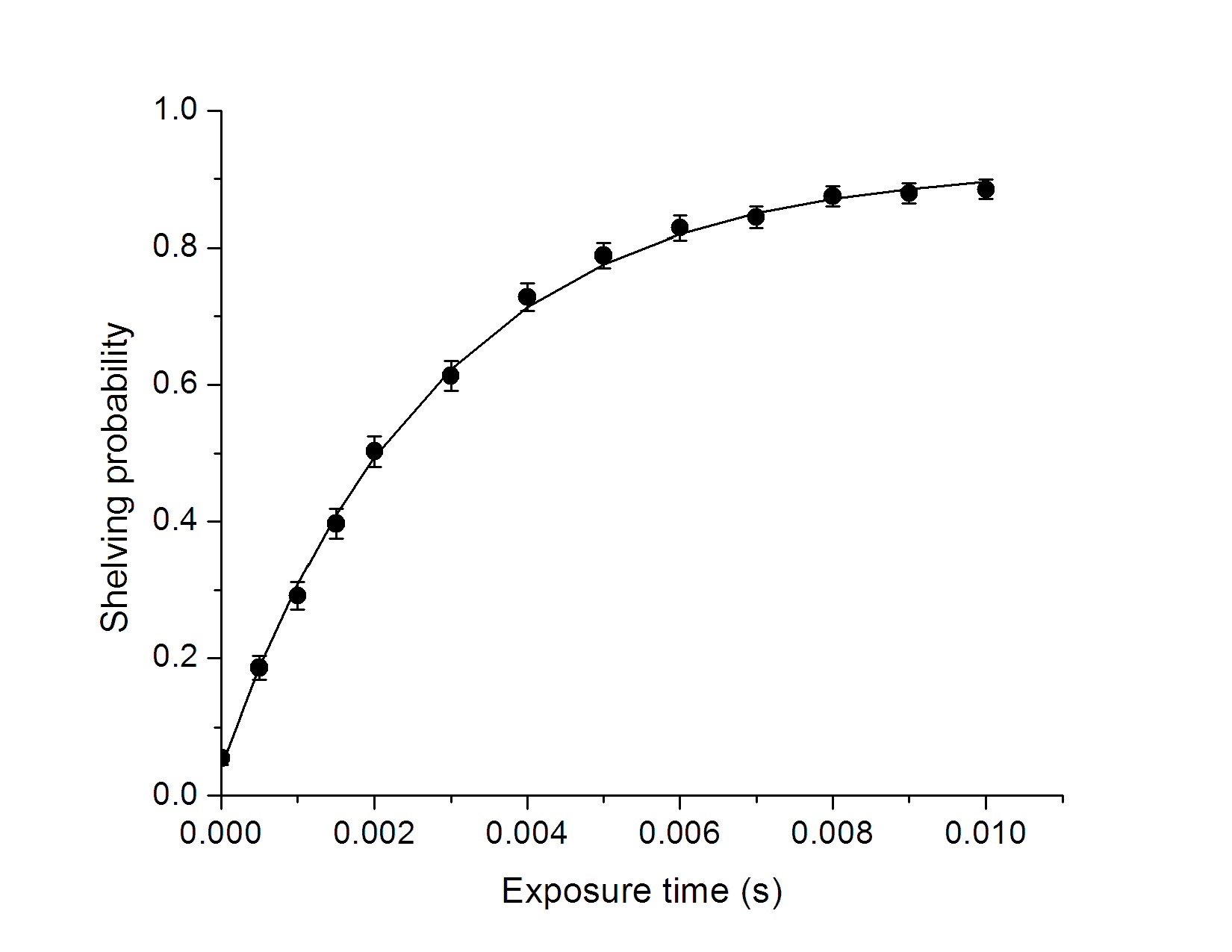}
\caption{Representative data obtained to establish a value for the relative strengths of the decays to the two D states in $^{138}\mathrm{Ba}^+$.  The ion was exposed to highly attenuated 455 nm light, weakly driving it to the $6\mathrm{P}_{3/2}$ state.  At long exposure times, the probability that the ion is dark will determine the relative proportion of the trials where the decay from $6\mathrm{P}_{3/2}$ was to $5\mathrm{D}_{5/2}$.  The solid line is a fit to the data using a three-parameter model, $P_{dark}=\Delta+B\left(1+e^{-\lambda t}\right)$, where the fit parameter $B$ determines the ratio of the relative strengths of the D state decays.}
\end{figure}

In order to establish absolute values for the decay ratios to all three states, one must devise a means to independently distinguish the $6\mathrm{S}_{1/2}$ and $5\mathrm{D}_{3/2}$ decays.  This is accomplished by driving the ground state to $6\mathrm{P}_{3/2}$ transition with single 400 fs pulses.  Similar excitation schemes have been employed on the $5\mathrm{S}_{1/2}\rightarrow5\mathrm{P}_{1/2}$ transition in atomic $^{85}\mathrm{Rb}$ \cite{Zhdanovich} and on the  $5\mathrm{S}_{1/2}\rightarrow5\mathrm{P}_{3/2}$ in $^{111}\mathrm{Cd}^+$ and $6\mathrm{S}_{1/2}\rightarrow6\mathrm{P}_{1/2}$ in $^{171}\mathrm{Yb}^+$ ions for use in coherent coupling of ion spin and photon polarization \cite{Madsen, Moehring}.  The pulses rotate the state of the ion through a Rabi angle $\theta$ determined by the energy and the length of the pulse.  The excited state then decays via the three available channels.  The $5\mathrm{D}_{5/2}$ state will be maximally populated at a Rabi angle $\theta$ equal to $\pi$.  By fitting values for the dark state probability at various values of the incident pulse energy with the expected form $P_{dark}=B^\prime\sin^2\left(\theta/2\right)$, where $\theta=\alpha\sqrt{E}$, the branching ratio $B^\prime$ for decays to $5\mathrm{D}_{5/2}$ can be extracted.  The fitting parameter corresponding to the excitation efficiency $\alpha$, a function of beam waist and position of the ion in the intensity profile of the beam, was found to be $3.41\pm0.07\:\mathrm{nJ}^{-1/2}$ and varied by no more than 10\% from run to run.  The fit parameter corresponding to the branching ratio $B^\prime$ proved robust to deviations in $\alpha$ at and beyond this level.  After multiple runs at decreasing and increasing attenuation (corresponding to increasing or decreasing the energy per pulse), data were binned by pulse energy and fit to the two-parameter model above, the results of which are displayed in Fig. 4.  Vertical error bars are statistical, corresponding to binomial standard error  in the average of all runs.  Systematic effects considered in this measurement include the effect of adjacent pulses, which was shown through numerical integration of the optical Bloch equations to have little effect on the maximum of the dark state probability, and the uncertainty in shot-to-shot pulse energy, found to be approximately 10\% which is indicated in the horizontal error bars.  Near the maximum, this pulse energy variation contributes to a systematic shift in the measured value of $B^\prime$ of -0.7 \%.  The bright ion counts resulting from improper synchronization of the pulse picking AOM with the continuous pulse train of the Ti:Sapphire oscillator were counted during the course of data collection and found to occur an average of $4\pm1$\% of the time and were removed from the data sets, leaving a 1\% systematic uncertainty.  The same systematics as in the previous measurement, namely the leakage of cooling light and collisions with background gas, again contribute approximately 0.5\% systematic uncertainty as previously mentioned.  The combined effect of these systematic effects leads to an overall systematic uncertainty of about 2\%.

\begin{figure}[h]
\centering
\includegraphics[scale=0.32]{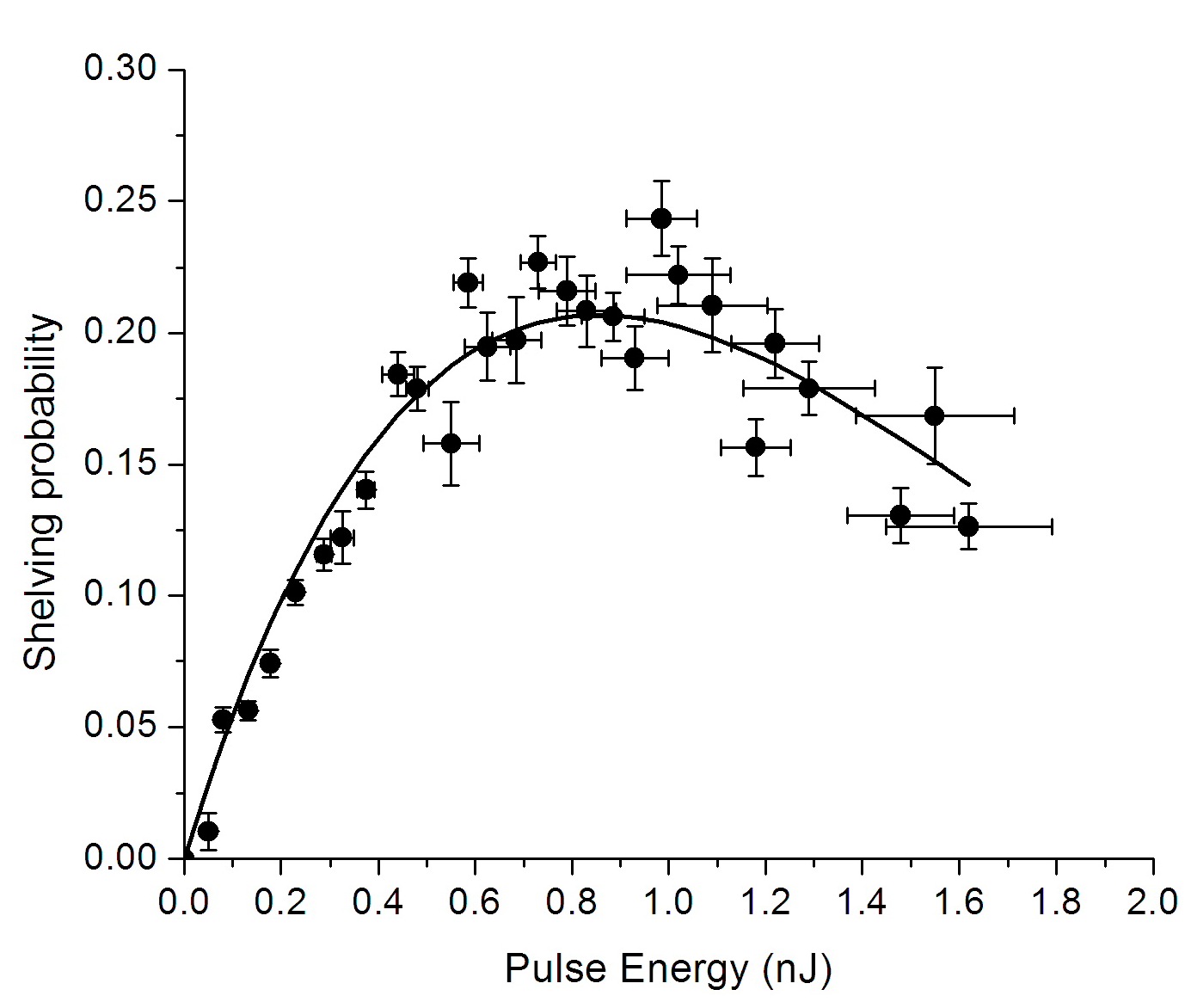}
\caption{Data obtained from single pulse excitation of the $6\mathrm{S}_{1/2}\rightarrow6\mathrm{P}_{3/2}$ transition.  Shelving probability represents the percentage of the runs where upon spontaneous decay, the ion was no longer fluorescing, i.\hspace{-6pt} e.\hspace{-6pt} was in the $5\mathrm{D}_{5/2}$ state.  The solid line is a two-parameter fit to the data using $P_{dark}=B^\prime\sin^2\left(\alpha\sqrt{E}/2\right)$, from which the value for the branching probability $B^\prime$ from $6\mathrm{P}_{3/2}$ to $5\mathrm{D}_{5/2}$ was extracted to be 0.215$\pm$0.0047.}
\end{figure}

Combining the information from the weak shelving and the Rabi rotation experiments, one can calculate absolute values for the three transition strengths.  Data are summarized in Table (1) and pictorially represented with overall error bars in Fig. 5, and compared to previously measured values \cite{Kastberg, Davidson, Gallagher, Reader} as well as calculated matrix elements \cite{Gopakumar, Guet, Dzuba}.  The results of our measurement represent an approximate threefold improvement over previously achieved accuracy for two of the three transitions, with the notable exception of the $5\mathrm{D}_{3/2}$ decay ratio whose error is approximately the same as the best previous measurement.  The  transition probability $A_{fi}$ for each decay can be calculated using an experimentally determined lifetime $\tau_{6P_{3/2}}=6.32(10)\:\mathrm{ns}$ \cite{Pinnington} and the fact that $\tau_i=\left(\sum_f A_{fi}\right)^{-1}$.  From these ratios, following the formal treatment of calculations of transition strengths \cite{Guet, Gopakumar}, the radial matrix elements for the involved transitions can be calculated.  We find $\left.\left<6\mathrm{P}_{3/2}\right|\right|\hat{D}_{E1}\left|\left|6\mathrm{S}_{1/2}\right>\right.=4.720\pm0.040$, $\left.\left<6\mathrm{P}_{3/2}\right|\right|\hat{D}_{E1}\left|\left|5\mathrm{D}_{3/2}\right>\right.= 1.349\pm0.036$ and $\left.\left<6\mathrm{P}_{3/2}\right|\right|\hat{D}_{E1}\left|\left|5\mathrm{D}_{5/2}\right>\right.=3.945\pm0.066$ in units where $ea_o=1$.  Our findings for the decay to the $6\mathrm{S}_{1/2}$ state show agreement with the calculated values of \cite{Gopakumar} to within one standard deviation and are within the error bars of the \cite{Reader} and \cite{Gallagher} measurements.  While our measurement of the $5\mathrm{D}_{3/2}$ decay rate did not improve the uncertainty in the measured value for that particular decay, it is characteristically lower than all previously measured values, though not outside of quoted error bars.  However, all calculated values for that particular decay fall within the bounds of our measurement.  Our value for the $5\mathrm{D}_{5/2}$ decay rate, which represents the most significant improvement in statistical uncertainty, sits at the lower bound of the \cite{Reader} and \cite{Gallagher} measurements.

\begin{table*}[h]
\caption{Calculation of transition probabilities and radial matrix elements from measured relative transition strengths from the $6\mathrm{P}_{3/2}$ state in BaII and comparison to previous work.
\label{Table 1}}
\begin{ruledtabular}
\begin{tabular}{cccccc}
Transition &\multicolumn{3}{c}{Present work} & Previous Experiment & Theory \\
 & Rel. Strength & D (a.u.) & $A_{fi}\times10^9\:\mathrm{s}^{-1}$ & $A_{fi}\times10^9\:\mathrm{s}^{-1}$ & D (a.u.) \\ \hline
 $6\mathrm{P}_{3/2}\rightarrow 6\mathrm{S}_{1/2}$ & 0.756$\pm$0.046 &  4.720$\pm$0.040 &  0.1196$\pm$0.0020 & \begin{tabular}{c}0.106$\pm$0.009\cite{Kastberg}\\0.117$\pm$0.004\cite{Reader}\\0.118$\pm$0.008\cite{Gallagher}\end{tabular} & \begin{tabular}{c}4.6982\cite{Gopakumar}\\4.658\cite{Guet}\\4.6738\cite{Dzuba}\end{tabular} \\ \hline
 $6\mathrm{P}_{3/2}\rightarrow 5\mathrm{D}_{3/2}$ & 0.0290$\pm$0.0015 &  1.349$\pm$0.036 &  0.004589$\pm$0.00025 & \begin{tabular}{c}0.00469$\pm$0.00029\cite{Kastberg}\\0.0048$\pm$0.0005\cite{Reader}\\0.0048$\pm$0.0006\cite{Gallagher}\end{tabular} & \begin{tabular}{c}1.2836\cite{Gopakumar}\\1.312\cite{Guet}\\1.2817\cite{Dzuba}\end{tabular} \\ \hline
 $6\mathrm{P}_{3/2}\rightarrow 5\mathrm{D}_{5/2}$ & 0.215$\pm$0.0064 &  3.945$\pm$0.066 & 0.03402$\pm$0.00115 & \begin{tabular}{c}0.0377$\pm$0.0024\cite{Kastberg}\\0.037$\pm$0.004\cite{Reader}\\0.037$\pm$0.004\cite{Gallagher}\end{tabular} & \begin{tabular}{c}3.9876\cite{Gopakumar}\\4.057\cite{Guet}\\3.8336\cite{Dzuba}\end{tabular} \\
\end{tabular}
\end{ruledtabular}
\end{table*}

Experimental confirmation of calculated values for these matrix elements is of particular use for several reasons.  The parity-nonconserving electric dipole transition amplitude for the transition $6\mathrm{S}_{1/2}\rightarrow5\mathrm{D}_{3/2}$ is given as a sum over matrix elements of both the dipole and weak charge Hamiltonian where the largest contributions to the sums come from the $6\mathrm{P}_{1/2}$ and $6\mathrm{P}_{3/2}$ terms, approximately 90 and 8\%, respectively \cite{Sahoo, Dzuba}.  The $\left.\left<6\mathrm{P}_{3/2}\right|\right|\hat{D}_{E1}\left|\left|5\mathrm{D}_{3/2}\right>\right.$ and $\left.\left<6\mathrm{P}_{3/2}\right|\right|\hat{D}_{E1}\left|\left|6\mathrm{S}_{1/2}\right>\right.$ terms appear explicitly in this amplitude.  Additional application of measured values for these matrix elements is of relevance to searches for drifts in the fine structure constant, both astronomically in anomalous quasar absorption spectra \cite{Berengut} and in Earth-based optical frequency atomic clocks \cite{clock}.

\begin{figure}[h]
 \centering
 \includegraphics[scale=0.35]{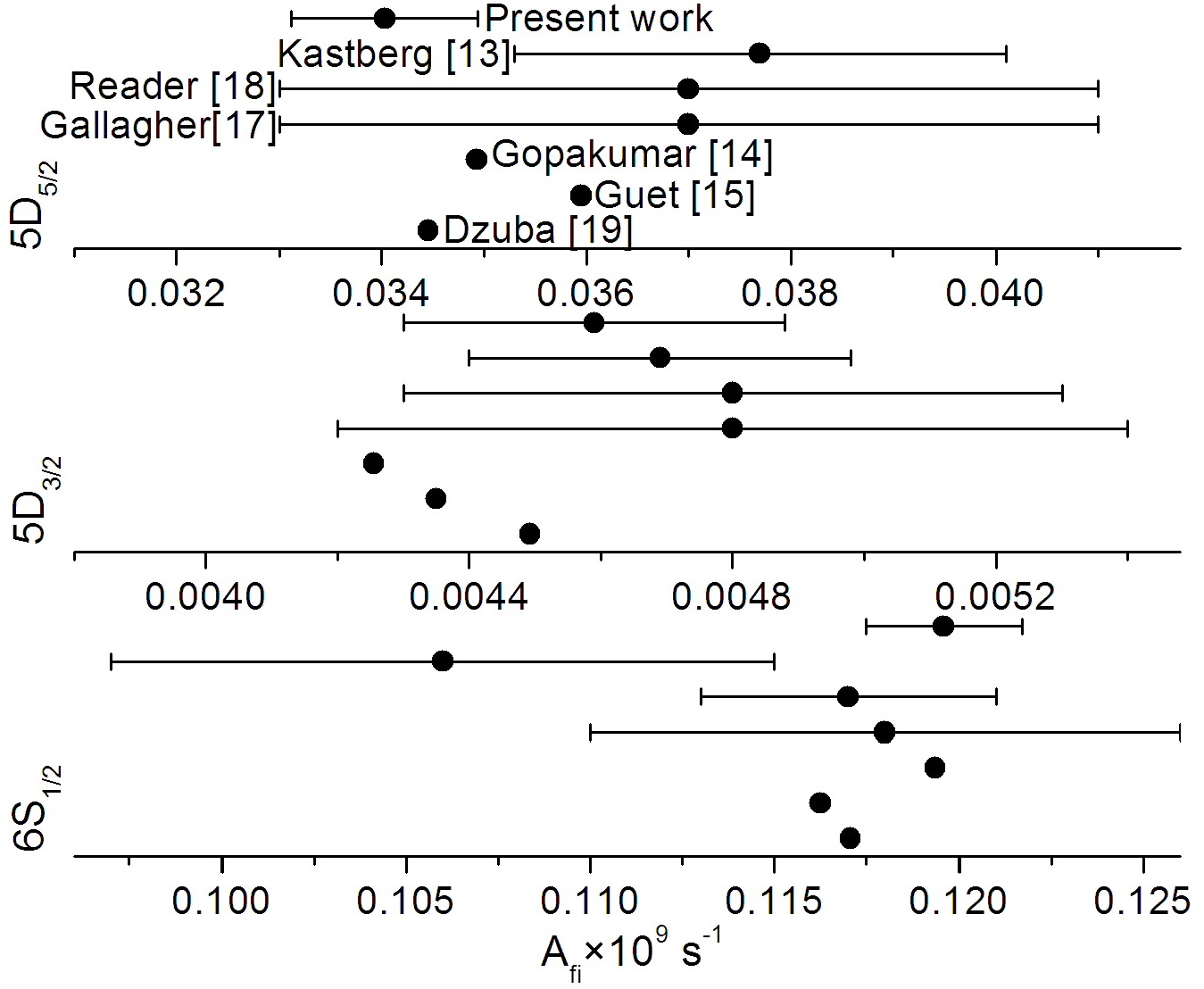}
 \caption{Comparison of presently measured transition rates for the three transitions from $6\mathrm{P}_{3/2}$ to the final state noted at the left vertical axis to previously measured and calculated values.  Error bars are one sigma confidence intervals on experimental data.  Vertical ordering in lower two panels is the same as topmost panel.}
 \end{figure}

In conclusion, we have measured the branching ratios from the $6\mathrm{P}_{3/2}$ state of $^{138}\mathrm{Ba}^+$ into all states allowed via dipole transitions using a single trapped ion excited with ultrafast optical pulses, and calculated transition probabilities and dipole matrix elements of the transitions.  For two of the three transitions, we achieve a threefold improvement in overall uncertainty, while for the $5\mathrm{D}_{3/2}$ decay we are approximately at the same level of precision as the best previous measurements.  The D state decays show a characteristically lower decay rate than previously measured and are closer to theoretical values.  Ultrafast excitation of ions using single femtosecond pulses has particular use for quantum computation when used with the 137 isotope of Ba$^+$.  Coherent coupling between the spin states of the ion and the photon polarization state has been demonstrated in $^{111}\mathrm{Cd}^+$ \cite{Madsen} and $^{171}\mathrm{Yb}^+$ \cite{Moehring} and its application to barium follows very closely the methods employed in the present work.

We would like to thank Norval Fortson and Jeff Sherman for helpful discussion and Adam Kleczewski, Li Wang, Gary Howell and Xiaoli Li for their helpful contributions to earlier stages of the experiment.  This research was supported by the University of Washington Royalty Research Fund, the Army Research Office DURIP grant and by the NSF AMO program.


\begin{thebibliography}{99}
\bibitem{CZ}J. I. Cirac \& P. Zoller, Phys. Rev. Lett. \bf 74\rm(20), 4091-4 (1995).
\bibitem{Monroe}C. Monroe, \it et. al., \rm Phys. Rev. Lett. \bf 75 \rm(25), 4714-7 (1995).
\bibitem{Berkeland}D. J. Berkeland \it et. al., \rm Phys. Rev. Lett. \bf80 \rm(10), 2089-92 (1998).
\bibitem{constants}V. A. Dzuba, V. V. Flambaum \& J. K. Webb, Phys. Rev. A \bf 59\rm, 230-7 (1999).
\bibitem{Fortson} N. Fortson, Phys. Rev. Lett. \bf 70\rm, 2383-6, (1993).
\bibitem{Koerber1}T. Koerber, M. Schacht, W. Nagourney \& E. Fortson, Jour. Phys. B \bf 36\rm, 637-48, (2003).
\bibitem{Koerber2}T. Koerber, \it et. al., \rm Phys. Rev. Lett. \bf 88\rm, 143002, (2002).
\bibitem{Pinnington}E. H. Pinnington, R. W. Berends \& M. Lumsden, J. Phys. B. \bf 28\rm, 2095 (1995).
\bibitem{Madej}A. A. Madej \& T. D. Sankey, Phys. Rev. A. \bf 41\rm(5), 2621-2630 (1990).
\bibitem{Zhdanovich}S. Zhdanovich, \it et. al., \rm Phys. Rev. Lett. \bf 100\rm, 103004 (2008).
\bibitem{Madsen}M. J. Madsen, \it et. al., \rm Phys. Rev. Lett. \bf 97\rm, 040505 (2006).
\bibitem{Moehring}D. L. Moehring, \it et. al., \rm Nature \bf 449\rm, 68-72 (2007).
\bibitem{Kastberg}A. Kastberg, \it et. al., \rm Jour. Opt. Soc. Am. B \bf 10\rm(8), 1330-6, (1993).
\bibitem{Gopakumar}G. Gopakumar, \it et. al., \rm Phys. Rev. A. \bf 66 \rm 032505-1-6, (2002).
\bibitem{Guet}C. Guet \& W. R. Johnson, Phys. Rev. A \bf 44\rm(3), 1531-1535 (1991).
\bibitem{Davidson}M. D. Davidson, \it et. al., \rm Astron. Astrophys. \bf 255\rm, 457-8, (1992).
\bibitem{Gallagher}A. Gallagher, Phys. Rev. \bf 157\rm, 24-30, (1967).
\bibitem{Reader}J. Reader, \it et. al., Wavelengths and transition probabilities for atoms and atomic ions\rm, Natl. Bur. Stand. Ref. Data Ser., Natl. Bur. Stand. (U.S.) Circ. No. 68 (GPO, Washington, D.C.), Vol. 10, (1980).
\bibitem{Dzuba}V. A. Dzuba, V. V. Flambaum \& J. S. M. Ginges, Phys. Rev. A \bf 63\rm, 062101 (2001).
\bibitem{Sahoo}B. K. Sahoo, \it et. al., \rm Phys. Rev. A \bf 68\rm, 040501 (2003).
\bibitem{Berengut}J. C. Berengut, V. A. Dzuba \& V. V. Flambaum, Phys. Rev. A \bf 68\rm, 022502-1-6 (2003).
\bibitem{clock}V. A. Dzuba \& V. V. Flambaum, Phys. Rev. A \bf 61\rm, 034502 (2000).
\bibitem{Yu}N. Yu, W. Nagourney, \& H. Dehmelt, Phys. Rev. Lett. \bf 78\rm, 4898 - 4901 (1997).
\end{thebibliography}
\end{document}